\documentclass[sigconf]{acmart}

\renewcommand\footnotetextcopyrightpermission[1]{}
\usepackage{array}
\newcolumntype{H}{>{\setbox0=\hbox\bgroup}c<{\egroup}@{}}
\usepackage{multirow}
\usepackage{caption, makecell}
\usepackage{url}
\usepackage{hyperref}
\usepackage{amsmath}
\usepackage{amsfonts}
\usepackage{dsfont}
\usepackage{relsize}
\usepackage{listings}
\usepackage{lipsum}
\usepackage{color}
\usepackage[inline]{enumitem}
\usepackage{algorithm}
\usepackage{algpseudocode}
\usepackage{wrapfig}
\usepackage{pifont}
\usepackage{graphicx}
\usepackage{dblfloatfix}
\usepackage{balance}
\usepackage[htt]{hyphenat}
\usepackage{pgfplotstable,booktabs}
\usepackage{color}
\definecolor{lightyellow}{rgb}{1,1,0.64}

\usepackage{graphicx}

\newboolean{showcomments}

\setcopyright{none}

\usepackage{booktabs}   
\usepackage{subcaption} 
\usepackage[T1]{fontenc}
\usepackage[latin1]{inputenc}

\newcommand{\Q}[1]{\textbf{RQ#1}}

\newcommand{\algosize}{\footnotesize}
\newcommand{\tablesize}{\footnotesize}

\newcommand{\tool}{InspectJS}
\newcommand{\inferenceengine}{$\text{Seldon}^{*}$}
\newcommand{\inferenceengineFunc}{$\text{Seldon}^{++}$}

\newcommand{\getsinks}{GetSinks}

\newcommand{\similarityrefiner}{Similarity-Based Refiner}
\newcommand{\similarityrefinerFunc}{SimilarityBasedRefiner}

\newcommand{\feedbackrefiner}{Feedback-Based Refiner}
\newcommand{\feedbackrefinerFunc}{FeedbackBasedRefiner}

\newcommand{\Space}[1]{}
\newcommand{\CComment}[1]{}

\newcommand{\mypara}[1]{\vspace{.03in}\noindent \textbf{#1.}}

\newcommand{\etal}{et al.}

\definecolor{ghgreen}{rgb}{0.90,1,0.93}
\definecolor{ghred}{rgb}{1,0.88,0.94}
\definecolor{ghblue}{rgb}{0.95,0.98,1.0}

\newcommand\dg[1]{\nbc{DG}{#1}{teal}}
\newcommand\ms[1]{\nbc{MS}{#1}{blue}}
\newcommand\resolved[1]{}

\newcommand{\nver}[2]{#1^{\textit{#2}}}
\newcommand{\ever}[2]{#1_{\textit{#2}}}

\newcommand{\code}[1]{\texttt{#1}}
\newcommand{\simm}{Z}

\definecolor{WowColor}{rgb}{.75,0,.75}
\definecolor{SubtleColor}{rgb}{0,0,.50}

\renewcommand{\Comment}[1]{}

\newcounter{margincounter}
\newcommand{\displaycounter}{{\arabic{margincounter}}}
\newcommand{\incdisplaycounter}{{\stepcounter{margincounter}\arabic{margincounter}}}

\let\origthelstnumber\thelstnumber
\makeatletter
\newcommand*\Suppressnumber{%
  \lst@AddToHook{OnNewLine}{%
    \let\thelstnumber\relax%
     \advance\c@lstnumber-\@ne\relax%
    }%
}

\newcommand*\Reactivatenumber[1]{%
  \setcounter{lstnumber}{\numexpr#1-1\relax}
  \lst@AddToHook{OnNewLine}{%
   \let\thelstnumber\origthelstnumber%
   \refstepcounter{lstnumber}
  }%
}

\ifthenelse{\boolean{showcomments}}
{
	\newcommand{\del}[1]{\textcolor{red}{\sout{#1}}}    
}{
	\newcommand{\del}[1]{}                              
	
}

\ifthenelse{\boolean{showcomments}}{
	\newcommand{\nbc}[3]{
		{\colorbox{#3}{\bfseries\sffamily\scriptsize\textcolor{white}{#1}}}
		{\textcolor{#3}{\sf\small$\langle$\textit{#2}$\rangle$}}}
}{
	\newcommand{\nbc}[3]{}
	
}

\ifthenelse{\boolean{showcomments}}{

\newcommand{\TBD}[1]{\textcolor{SubtleColor}{ {\tiny \bf (!)} #1}}

\newcommand{\fTBD}[1]{\textcolor{SubtleColor}{$\,^{(\incdisplaycounter)}$}\marginpar{\tiny\textcolor{SubtleColor}{ {\tiny $(\displaycounter)$} #1}}}

}{

\newcommand{\fTBD}[1]{}
\newcommand{\TBD}[1]{}

}

\definecolor{lightgray}{rgb}{.9,.9,.9}
\definecolor{darkgray}{rgb}{.4,.4,.4}
\definecolor{purple}{rgb}{0.65, 0.12, 0.82}

\lstdefinelanguage{JavaScript}{
	keywords={typeof, new, true, false, catch, function, return, null, catch, switch, var, if, in, while, do, else, case, break, await, let, const, async, try},
	keywordstyle=\color{blue}\bfseries,
	ndkeywords={class, export, boolean, throw, implements, import, this},
	ndkeywordstyle=\color{darkgray}\bfseries,
	identifierstyle=\color{black},
	sensitive=false,
	comment=[l]{//},
	morecomment=[s]{/*}{*/},
	commentstyle=\color{purple}\ttfamily,
	stringstyle=\color{red}\ttfamily,
	morestring=[b]',
	morestring=[b]",
	morestring=[b]`
}

\begin{document}

\title{InspectJS: Leveraging Code Similarity and User-Feedback for
  Effective Taint Specification Inference for JavaScript}


\author{Saikat Dutta}
\affiliation{
  \institution{UIUC}            
  \city{Urbana}
  \country{USA}                    
}
\email{saikatd2@illinois.edu}          

\author{Diego Garbervetsky}
\affiliation{
	\institution{DC/UBA. ICC/CONICET}
	\city{Buenos Aires}
	\country{Argentina}
}
\email{diegog@dc.uba.ar}

\author{Shuvendu Lahiri}
\affiliation{
	\institution{Microsoft Research}
	\city{Seattle}
	\country{USA}
}
\email{shuvendu.lahiri@microsoft.com}

\author{Max Sch\"afer}
\affiliation{
  \institution{GitHub}
  \city{Oxford}
  \country{UK}
}
\email{max-schaefer@github.com}

\begin{abstract}
Static analysis has established itself as a weapon of choice for detecting
security vulnerabilities. Taint analysis in particular is a very general and
powerful technique, where security policies are expressed in terms of forbidden
flows, either from untrusted input sources to sensitive sinks (in integrity
policies) or from sensitive sources to untrusted sinks (in confidentiality
policies). The appeal of this approach is that the taint-tracking mechanism has
to be implemented only once, and can then be parameterized with different taint
specifications (that is, sets of sources and sinks, as well as any sanitizers
that render otherwise problematic flows innocuous) to detect many different
kinds of vulnerabilities.

But while techniques for implementing scalable inter-procedural static taint
tracking are fairly well established, crafting taint specifications is still
more of an art than a science, and in practice tends to involve a lot of manual
effort.

Past work has focussed on automated techniques for inferring taint
specifications for libraries either from their implementation or from the way
they tend to be used in client code. Among the latter, machine learning-based
approaches have shown great promise.

In this work we present our experience combining an existing machine-learning
approach to mining sink specifications for JavaScript libraries with manual
taint modelling in the context of GitHub's CodeQL analysis framework. We show
that the machine-learning component can successfully infer many new taint sinks
that either are not part of the manual modelling or are not detected due to
analysis incompleteness. Moreover, we present techniques for organizing sink
predictions using automated ranking and code-similarity metrics that allow an
analysis engineer to efficiently sift through large numbers of predictions to
identify true positives.
\end{abstract}

\maketitle

\section{Introduction}
It is a truth universally acknowledged, that a static analyzer in possession of
an inter-procedural taint analysis must be in want of taint specifications. Even
the most scalable taint analysis cannot, in general, cope with the vast amount
of third-party library code that even very simple modern software depends on,
quite apart from the fact that this code may be written in an entirely different
language (as is the case for native library bindings in scripting languages) or
may not even be available at all (for binary dependencies).

Taint specifications distill out the analysis-relevant information for such
libraries in a compact and reusable form. Specifically, a taint analysis is
usually interested in \emph{source specifications}, indicating library APIs that
may return untrusted (``tainted'') data possibly controlled by a malicious
attacker, and \emph{sink specifications}, identifying APIs into which such
tainted data must not flow without appropriate sanitization, which is in turn
captured by \emph{sanitizer specifications}. Other potentially interesting
specifications include propagation specifications modelling whether a function
propagates taint from its arguments to its return value (a dual to sanitizer
specifications), aliasing specification modelling any aliasing relationships
introduced by the function, and others.

In practice, these specifications are often manually crafted by analysis
engineers based on library documentation or code. While this allows maximum
flexibility and precision, it is a labor-intensive and error-prone process,
often leading to missing or spurious models, which in turn cause missing or
spurious analysis alerts.

\begin{figure*}
  \begin{tabular}{cc}
\begin{lstlisting}[language=JavaScript]
sliderController.SaveSlider = async (req, res, nxt) => {
  try {
    const slider = req.body/*#\ding{192}#*/; /*#\label{line:eg1:slider} #*/
    let id = slugify(slider.slider_key); /*#\label{line:eg1:slugify} #*/
    await sliders.findByIdAndUpdate({ id: id/*#\ding{194}#*/ },  /*#\label{line:eg1:updstart} #*/
    {
      $set: slider,
    }); /*#\label{line:eg1:updend} #*/
    ...
  } catch (err) {
    nxt(err);
  }
};

function slugify(text) {  /*#\label{line:eg1:slugifystart} #*/
  return text.toLowerCase().replace(/\s+/g, '-')/*#\ding{193}#*/; /*#\label{line:eg1:replace} #*/
} /*#\label{line:eg1:slugifyend} #*/
\end{lstlisting}
&
\begin{lstlisting}[language=JavaScript]
loginlogController.logout = async (req, res, nxt) => {
  try {
    let token = req.body.token/*#\ding{195}#*/;  /*#\label{line:eg2:token} #*/
    token = token.replace('Bearer ', '')/*#\ding{196}#*/; /*#\label{line:eg2:replace} #*/
    await loginlogs.findOneAndUpdate({ token: token/*#\ding{197}#*/ }, /*#\label{line:eg2:callstart} #*/
    {
      $set: { is_active: false, logout_date: Date.now() }
    }); /*#\label{line:eg2:callend} #*/
    console.log(token/*#\ding{198}#*/); /*#\label{line:eg2:log} #*/
    ...
  } catch (err) {
    nxt(err);
  }
};
\end{lstlisting} \\
  (a) & (b)
  \end{tabular}
 \vspace{-1em} 
  \caption{Two uses of APIs relevant to NoSQL injection vulnerabilities: (a)
  \code{findByIdAndUpdate}, and (b) \code{findOneAndUpdate}. Circled numbers
  indicate expressions referenced in the text.}
  \label{fig:motivating-example}
  \vspace{-1em}
\end{figure*}

Many different techniques have been proposed in the literature to instead
generate taint specifications automatically, either from the source code of the
library or from examples of its usage. The former typically involves some sort
of summarization analysis being done on the library source code. Our approach is
based on Seldon~\cite{seldon}, a representative of the latter category, which
works by mining a (large) corpus of client code for the library in question, and
then uses probabilistic inference to identify candidate taint specifications
from the way that code interacts with the library. The inference attaches to
each candidate a score between zero and one which intuitively indicates how
confident we are that the prediction is correct. As a final step, the concrete
candidates identified on the training set need to be abstracted into a code-base
independent representation that can be used to find candidate taint
specifications on other code bases.

In this work, we add to this process a refinement step where the score of a
candidate is adjusted using code-similarity metrics, giving greater weight to
candidates that appear in a context that is syntactically similar to known taint
specifications for which we already possess a manually-written model.

Unlike Seldon, our goal is not to obviate manual modelling but instead to use
automated specification mining as a driver for detecting missing or incomplete
models. To make this feasible, we need a way of presenting predicted taint
specifications to an analysis engineer that makes them easy to triage and
efficiently prune away false positives. 

We propose three criteria for organizing predictions: by their score (as
determined by the probabilistic inference and refined using code similarity), by
their generality, and by their similarity to each other. The first one is quite
obvious: predictions with a low score are not worth showing to the engineer. For
the second one, the idea is that overly general representations that lead to a
large number of predicted sinks are unlikely to be true positives. The third one
again uses code similarity, this time to allow the engineer to dismiss a false
positive along with all other predictions that are syntactically similar to it.

We have implemented our approach in a tool called \tool{}, which is based on the
CodeQL analysis framework~\cite{codeql}, and can be used to infer sink
specifications for JavaScript.

We motivate our work using a concrete example in
Section~\ref{sec:motivating-example}, discuss its relationship with Seldon in
Sections~\ref{sec:seldon} and~\ref{sec:approach}, and empirically evaluate
the quality of the sink predictions produced by \tool{} in
Section~\ref{sec:evaluation} before surveying related work in
Section~\ref{sec:related-work} and concluding in Section~\ref{sec:conclusion}.

In summary, the main contributions of our work are:

\begin{itemize}
	\item A novel combination of a probabilistic approach to predicting
	taint-sink specifications from static data-flow information with
	code-similarity based refinement to adjust prediction score based on their
	similarity to known sinks.
	\item Three techniques for organizing sink predictions based on their score,
	generality, and similarity to each other, allowing a domain expert to
	efficiently triage large numbers of automatically generated predictions.
	\item An implementation of our technique on top of the CodeQL static analysis
	framework in a tool called \tool{}.
	\item An empirical evaluation of the quality of the predictions produced by
	\tool{} on real-world code bases, showing that it correctly identifies taint
	sinks that are missing from the manually-written models shipping with
	CodeQL. We have reported some of these to the CodeQL library maintainers,
	and they have incorporated our suggestions into the models.
\end{itemize}

\section{Motivating example}\label{sec:motivating-example}
To motivate our work, we will show an example of a missing taint
specification in the CodeQL static analysis for JavaScript, which was
identified with the help of \tool{} and has since been added to the
manually-written
model.\footnote{\url{https://github.com/github/codeql/pull/4753}}

Consider the code snippet in Figure~\ref{fig:motivating-example}(a),
which is adapted (and slightly simplified) from the WaftEngine
project.\footnote{\url{https://github.com/WaftTech/WaftEngine}} It
shows a route handler from an HTTP server implemented as a JavaScript
function accepting three parameters \code{req}, \code{res}, and
\code{nxt}. Parameter \code{req} is the HTTP request object
originating from a client, \code{res} is the response object to be
filled in by the handler, and \code{nxt} is the next handler to be
called in case of error.

The route handler extracts the \code{slider\_key} field from the
request body (Line~\ref{line:eg1:slider}), passes it to the
\code{slugify} function (defined in
Lines~\ref{line:eg1:slugifystart}-\ref{line:eg1:slugifyend}) which
lower-cases it and replaces all spaces with hyphens
(Line~\ref{line:eg1:slugify}), and then uses the resulting string to
look up and update an entry in a NoSQL database using the
\code{findByIdAndUpdate} method
(Lines~\ref{line:eg1:updstart}-\ref{line:eg1:updend}). The first
argument to this method is a JavaScript object, which is interpreted
as a NoSQL query. For example, the query \code{\{ id: "myslider" \}}
selects all entries with the \code{id} field equal to
\code{"myslider"}. This query is really a short-hand for the query
\code{\{ id: \{ \$eq: "myslider" \} \}}, using the MySQL operator
\code{\$eq} to compare the field \code{id} with the value
\code{"myslider"}.

Other operators allow more complicated tests, for example \code{\$ne}
for inequality, \code{\$regex} for regular expression matching, and
\code{\$where} for specifying an arbitrary JavaScript expression. It
is because of these more advanced operators that it is not, in
general, safe to pass data controlled by an untrusted user to a NoSQL
API method expecting a query, since the user might specify a query
using \code{\$ne} or \code{\$regex} to access almost any entry, or a
query using \code{\$where} to execute arbitrary JavaScript
code.\footnote{In practice, the code will usually be executed in a
	sandbox curtailing access to sensitive resources, but a malicious
	user could still specify a non-terminating condition to mount a
	denial-of-service attack.}

In this example, while \code{req.slider\_key} is under user control,
it is used in a reasonably safe manner: as revealed by its use in the
\code{slugify} function, \code{slider\_key} is a string, so it cannot
be used to encode potentially problematic conditions.  CodeQL
recognizes this and does not flag this snippet as problematic: while
its models allow it to classify \code{req.body} and its properties as
taint sources and the first argument to \code{findByIdAndUpdate} as a
taint sink, it also knows that
\code{replace}~(Line~\ref{line:eg1:replace}) acts as a sanitizer in
this case since its result is guaranteed to be a string.

Now consider the code snippet in
Figure~\ref{fig:motivating-example}(b), showing a different handler
function in the same project. Its structure is very similar: a
property of \code{req.body} is read (Line~\ref{line:eg2:token}),
processed with \code{replace} (Line~\ref{line:eg2:replace}), and then
used in a NoSQL query
(Lines~\ref{line:eg2:callstart}-\ref{line:eg2:callend}), this time
with the \code{findOneAndUpdate} method. Prior to our work CodeQL did
not recognize this method as a sink, and hence would have failed to
flag\fTBD{does CodeQL flag a safe use? in this case i guess it would
	just do nothing? \ms{I'm not sure I understand the question. Why would CodeQL flag a safe use?}}
not just this safe use, but also unsafe uses.

This is not an uncommon problem: manually modelling large APIs, like
the Mongoose framework\footnote{\url{https://mongoosejs.com/}} being
used here, is tedious and error-prone often leading to missing 
taint sources or sinks. Automated taint-specification mining promises
to eliminate or at least alleviate this problem.

Many different approaches have been proposed in the literature for
automatically discovering taint-specifications. Our work builds on the
\emph{flow triple} approach introduced by
Merlin~\cite{livshits2009merlin} as refined by
Seldon~\cite{seldon}. At a high level, this involves three steps.

First, we mine a training set of code bases for triples of program elements
$\langle\textit{src}, \textit{san}, \textit{snk}\rangle$ where taint may
propagate from \textit{src} to \textit{snk} via \textit{san}, and \textit{src}
is of a syntactic structure that means it could potentially be a taint source
(e.g., the result of a function call or a parameter to a callback), \textit{san}
could be a sanitizer (i.e., a function call), and \textit{snk} could be a taint
sink (e.g., a function argument). Note that this is done based on purely
syntactic criteria, independent of any semantic modelling.

For the snippets in Figure~\ref{fig:motivating-example}, for example, we would
obtain the triples
$\langle\text{\ding{192}},\text{\ding{193}},\text{\ding{194}}\rangle$ and
$\langle\text{\ding{195}},\text{\ding{196}},\text{\ding{197}}\rangle$
representing the flow from the request objects through the sanitizing string
replacements into the NoSQL queries.

Second, we perform a probabilistic analysis of these triples based on
the following observation: if \textit{src} is known to be a taint
source and \textit{san} a sanitizer, then it is very likely that among
all the nodes $\textit{snk}_i$ for which we have observed a flow
triple $\langle\textit{src}, \textit{san}, \textit{snk}_i\rangle$, at
least one is a taint sink, since otherwise there presumably will not
be any need for sanitization. Similarly, from known sources and sinks
we can infer the presence of a sanitizer, and from known sanitizers
and sinks a source.

For example, for the triple
$\langle\text{\ding{195}},\text{\ding{196}},\text{\ding{197}}\rangle$ mined from
Figure~\ref{fig:motivating-example}(b), we already know that \ding{195} is a
source and \ding{196} is a sanitizer, suggesting that \ding{197} may be a sink,
as is indeed the case.

These newly inferred elements can then be plugged into the triples in turn,
allowing us to discover even more sources, sinks, and sanitizers. As discussed
in more detail below, Seldon associates a score between 0 and 1 with each such
prediction which represents the degree of confidence in the correctness of the
prediction.

In the third step, we can suitably abstract the concrete elements observed on
the training set into a code-base independent representation, discard
predictions with low scores, and then use them to improve (``boost'') a taint
analysis, allowing it to flag more alerts on any code base, not just the ones in
the training set.

Alternatively (and this is the use case we are most interested in) the results
of the inference step can be presented to an analysis engineer for further
triaging, allowing them to identify lacunae in the hand-written models and
improve the analysis accordingly.

One weakness of such purely probabilistic approaches is that they
have little built-in knowledge of the semantics of the code being
analyzed apart from information about known taint specifications,
which can sometimes lead to surprising mispredictions.

For example, the call to \code{console.log} in
Figure~\ref{fig:motivating-example}(b) is not a sink, but based on
purely syntactic criteria it looks like a plausible candidate, and so
we would add the
$\langle\text{\ding{195}},\text{\ding{196}},\text{\ding{198}}\rangle$
to our set of mined flow triples. If there are sufficiently many
similar usages, we might then end up wrongly predicting that
\code{console.log} is a sink.

To prevent this, our approach combines the probabilistic analysis with
a post-processing step based on code similarity, whereby the scores of
sink predictions are adjusted based on their similarity to known
sinks. In our example, the call to \code{console.log} does not look
similar to a sink, so its score would be decreased, while the call to
\code{findOneAndUpdate} is syntactically quite similar to the call to
\code{findByIdAndUpdate}, which we know to be a sink, causing its
score to be increased.
\section{Background: Seldon}\label{sec:seldon}

Before we delve into the technical details of our approach, we give a
brief overview of Seldon~\cite{seldon}, on which the core inference
engine in \tool{} is based.

Seldon is a semi-supervised approach for inferring likely taint specifications
(source, sanitizer, or sink) for unmodeled or partially modeled library APIs
from a large corpus of client code using these APIs. Based on a set of client
programs $\mathcal{D}$ with a (small) set of program elements $\mathcal{A}_M$
already annotated as sources, sanitizers, or sinks, Seldon infers specifications
for the remaining (larger) set of un-annotated program elements $\mathcal{A}_U$.
This involves four steps described in more detail below: capturing information
flow in the form of a \emph{propagation graph}; representing the nodes of that
graph in a code-base independent form; building a constraint system encoding the
taint-specification inference problem; and finally solving that system.

While Seldon was originally implemented and evaluated for Python, we
adapt the approach for JavaScript as we describe in
Section~\ref{sec:adaptseldon}.

\mypara{Capturing information flow} For each input program, Seldon builds a
propagation graph $G=(V, E)$ where the edges $E$ capture information flows
between program elements $V$ (referred to as ``events'' in the original Seldon
paper, a usage which we will not follow). Program elements represented in the
propagation graph include arguments to and return values of function calls,
reads and writes of object properties or global variables, and any other
construct that propagates information. Seldon uses standard points-to analysis
to build such graphs.

\mypara{Representing program elements} By their very nature, program
elements are specific to a single code base, so in order to make sink
predictions reusable across programs we need to assign code-base
independent representations $\textsc{Rep}(v)$ to program elements
$v\in V$. Seldon uses a variant of qualified names for this purpose;
for example, the representation of the result of a function call could
be the fully qualified name of the function (most specific), or an
unqualified name (least specific).  For example, the method
\code{findOneAndUpdate} used in Figure~\ref{fig:motivating-example}(b)
has the two representations \code{mongoose.Model.findOneAndUpdate} and
\code{findOneAndUpdate}.

\mypara{Building a constraint system} Seldon frames the problem of
inferring taint specifications as a linear optimization task which can
be solved using efficient solvers and makes their solution scalable.

For each program element $v$ in $V$ and each representation
$n\in\textsc{Rep}(v)$ of $v$, Seldon instantiates three variables:
$\nver{n}{src}$, $\nver{n}{san}$, and $\nver{n}{snk}$, each of which denotes the
likelihood of $v$ being a source, sanitizer, or sink. Seldon adds constraints
for: ($1$) constraining each variable to $[0,1]$ to enable interpreting them as
probabilities, and ($2$) setting appropriate variables to $1$ or $0$ for program
elements whose specifications are known (from $\mathcal{A}_M$).

\begin{figure}[!thb]
  \centering
  \begin{subfigure}{0.4\linewidth}
    \centering
    \includegraphics[scale=0.2]{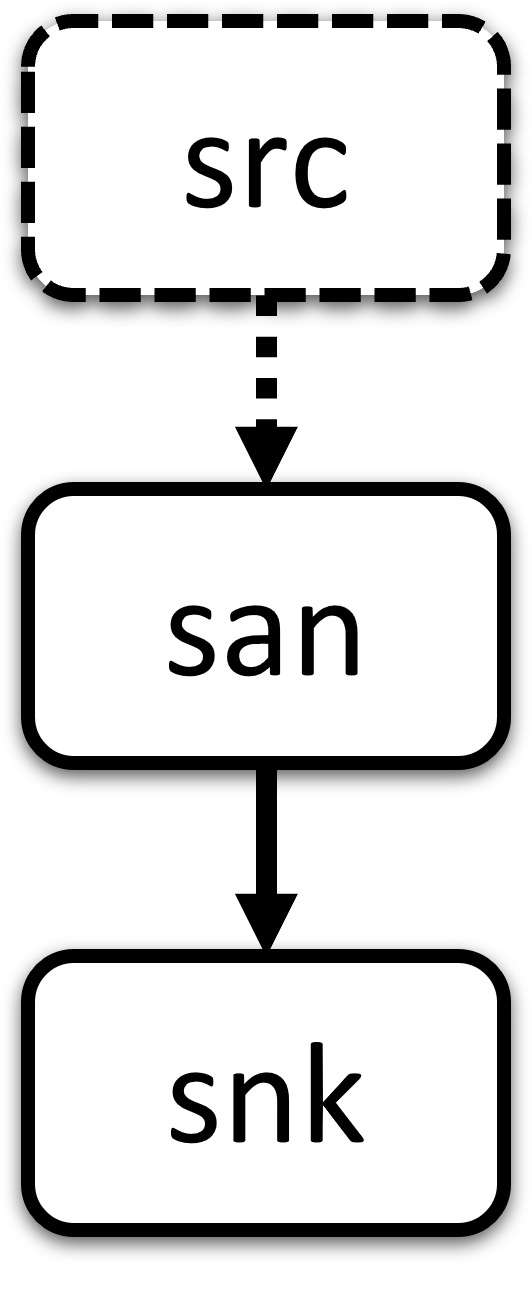}
  \vspace{-.5em}
    \caption{}
    \label{fig:intuition}
  \end{subfigure}
  \begin{subfigure}{0.4\linewidth}
    \centering
    \includegraphics[scale=0.2]{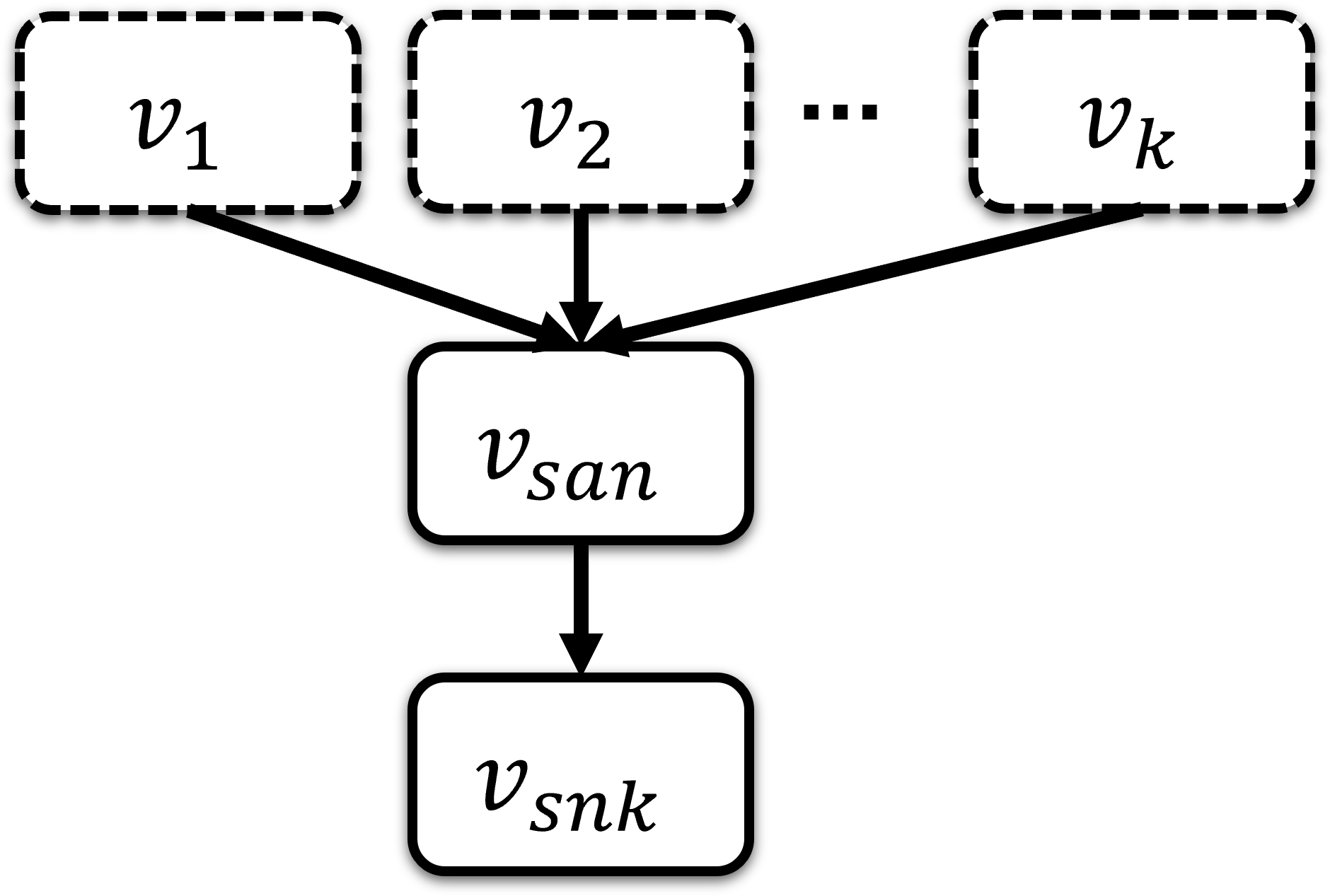}
  \vspace{-.5em}
    \caption{}
    \label{fig:intuitionevents}
  \end{subfigure}
  \begin{subfigure}{0.5\linewidth}
    \begin{equation*}    
		\nver{(\ever{n}{san})}{san} + \nver{(\ever{n}{snk})}{snk} \le
		\mathlarger{\sum}_{i=1}^{k}{\nver{(\ever{n}{i})}{src}}
                +\quad C \quad+ \quad\epsilon
    \end{equation*}
  \vspace{-1em}
    \caption{}
    \label{fig:intuitioneq}
  \end{subfigure}
  \vspace{-1em}
  \caption{Intuition of Information Flow Constraints}
  \label{fig:flowcons}
  \vspace{-0.1in}
\end{figure}

Seldon then adds various constraints to encode their intuitions about
information flow using the propagation
graphs. Figure~\ref{fig:flowcons} presents a visualization of one such
constraint. Figure~\ref{fig:intuition} indicates that if there is a
flow from a sanitizer to a sink, then it is most likely sanitizing the
output from a source. Figure~\ref{fig:intuitionevents} shows a
propagation graph capturing such an occurrence. It indicates that
if we have a program element $\ever{v}{san}$, classified as sanitizer, and
another program element $\ever{v}{snk}$, classified as sink, and there is a flow from
$\ever{v}{san}$ to $\ever{v}{snk}$, then we must classify at least one
of the program elements $v_i, i \in [1,\ldots,k]$ which flow into
$\ever{v}{san}$ as a \emph{source}. Figure~\ref{fig:intuitioneq}
presents the corresponding constraint that we add to encode this
intuition.  

Here, $\ever{n}{san}=\textsc{Rep}(\ever{v}{san})$,
$\ever{n}{snk}=\textsc{Rep}(\ever{v}{snk})$,
$\ever{n}{i}=\textsc{Rep}(\ever{v}{i})$. $C$ is a fixed
constant. Since the programs in the dataset may not always strictly
respect the intuitions, Seldon also adds the $\epsilon$ variables (one
for each constraint) to allow for minor deviations from the
assumptions. Seldon adds one such constraint for each $(\textit{san},
\textit{snk})$ pair which has at least one source candidate flowing
into the sanitizer. Seldon also adds other constraints analogously for
pairs of $(\textit{src}, \textit{snk})$ and $(\textit{src},
\textit{san})$.

\mypara{Solving the constraint system} Finally, Seldon solves the
optimization problem by minimizing the sum of all relaxation variables
($\epsilon_i$) and the sum of all variable scores: $\nver{n}{t}$, for
$t \in \{\textit{src},\textit{san},\textit{snk}\}$, subject to the
specified constraints. Seldon then returns the confidence scores for
each representation being a source, sink, or sanitizer. The inferred
specifications (e.g., with some minimum confidence) can then be used
to boost a taint analyzer. The spec is given by a list of triples $(n,
kind, score)$ where $kind \in \{src,snk, san\}$ and $n$ is a
representation.  Note that there can be many program elements with the
same representation.

\section{Our Approach}\label{sec:approach}
\begin{figure*}[!thb]
  \includegraphics[scale=0.5]{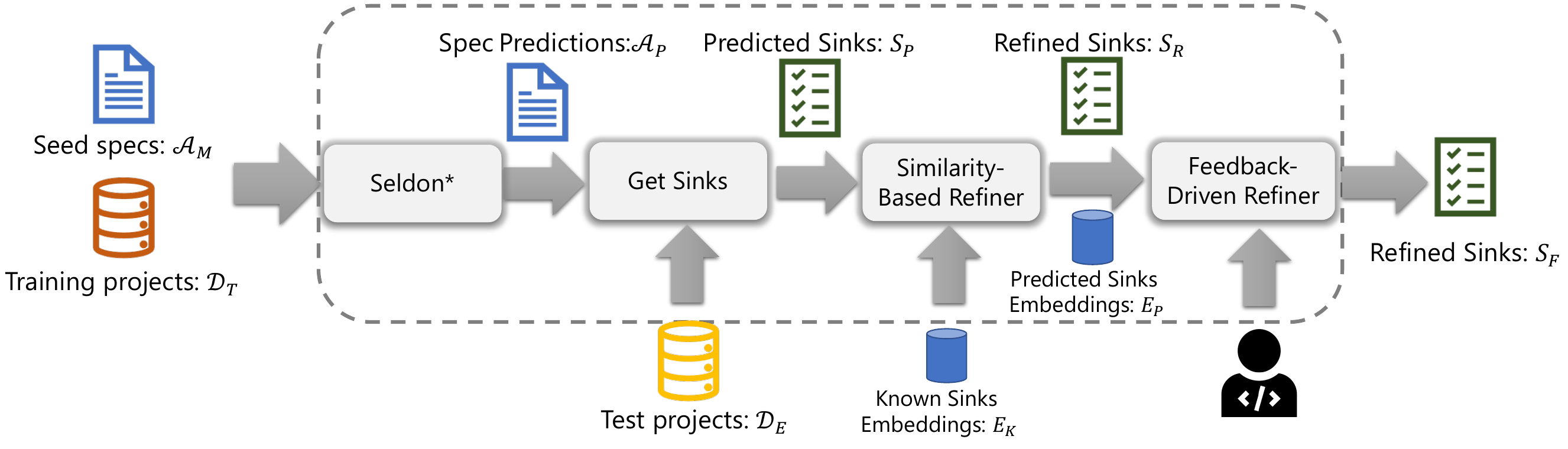}
  \vspace{-1em}
  \caption{\tool{}: System Overview}\label{fig:architecture}
  \vspace{-1em}
\end{figure*}

We now describe the key technical components of \tool{}.  At a high
level, \tool{} takes as input a training set of JavaScript code bases
$\mathcal{D}_T$ and a set of \emph{seed specifications}
$\mathcal{A}_M$ that can be used to identify \emph{known sinks}
$\mathcal{S}_M$, i.e., program elements in $\mathcal{D}_T$ that are
known to be sinks, for example, as the result of manual modeling.
\tool{} then processes the training set and infers a set of
\emph{predicted sinks}, i.e., program elements in $\mathcal{D}_T$ that
are likely to be sinks, but were previously unmodeled. Each of these
is associated with a \emph{score} between zero and one indicating the
likelihood that the element is, in fact, a sink. As the final step of
training, the predicted sinks, which are concrete program elements in
$\mathcal{D}_T$, are abstracted into code-base independent
\emph{representations}, which together with their associated scores
form the \emph{predicted sink specifications} $\mathcal{A}_P$, or
\emph{predictions} for short.

For a given test set of JavaScript code bases $\mathcal{D}_E$, these
predictions can be instantiated to yield new predicted sinks on those
code bases. Since some of the sinks may be false positives, they are
not intended to be used directly, but to go through a two-step
refinement process: one automated and one based on feedback from an
analysis engineer. The final, reviewed set of sinks can then be used
to improve manually-written models, or directly to find security
vulnerabilities.

\Comment{
\begin{algorithm}
  \caption{\tool{} Main Algorithm}\label{algo:main}
  \algosize{}
  \begin{flushleft}
    \hskip\algorithmicindent \textbf{Inputs}: Seed Specification $\mathcal{A}_M$, Training dataset $\mathcal{D}_T$, Evaluation dataset $\mathcal{D}_E$	 \\
    \hskip\algorithmicindent \textbf{Outputs}: Final Sinks $\mathcal{S}_F$
  \end{flushleft}
  \begin{algorithmic}[1]
    \Procedure{\tool{}}{$\mathcal{A}_M,  \mathcal{D}_T, \mathcal{D}_E$}
    \State $\mathcal{S}_U = \textsc{\inferenceengineFunc}(\mathcal{A}_M,  \mathcal{D}_T, \mathcal{D}_E)$
    \State $\mathcal{S}_R =
    \textsc{\similarityrefinerFunc}(\mathcal{S}_U, \mathcal{A}_M,
    \mathcal{D}_T)$
    \State $\mathcal{S}_F = \textsc{\feedbackrefinerFunc}(\mathcal{S}_R)$
    \State \Return $\mathcal{S}_F$
    \EndProcedure
  \end{algorithmic}
\end{algorithm}
}

Figure~\ref{fig:architecture} presents the overall architecture of
\tool{}.  These tasks are carried out by a pipeline of four
components: \textit{\inferenceengine{}}, \textit{\getsinks{}},
\textit{\similarityrefiner{}}, and \textit{\feedbackrefiner{}}.

\emph{\inferenceengine{}} implements our adaptation of Seldon's approach for
predicting likely sinks. It takes $\mathcal{A}_M$ and $\mathcal{D}_T$
as inputs and produces the predicted sink specifications
$\mathcal{A}_P$.  In general, a sink specification is a 3-tuple of the
form $\langle \textit{rep}, \textit{kind}, p \rangle$, where
$\textit{rep}$ is a program element representation, $\textit{kind} \in
\{\texttt{src}, \texttt{san}, \texttt{snk}\}$ denotes the role of the
program element, and $p \in [0,1]$ is a confidence score indicating
the likelihood of the element assuming the given role. In
$\mathcal{A}_M$, all representations are assigned a confidence score
of 1 (highest) since we already know their true roles.

\emph{\getsinks{}} then instantiates the predicted sinks
specifications $\mathcal{A}_P$ on test projects $\mathcal{D}_E$ to
produce new sinks, $\mathcal{S}_P$, which are tuples of the form
$\langle v, p \rangle$. Here, $v$ is a program element from one of the
test projects in $\mathcal{D}_E$ and $p \in [0, 1]$ is the confidence
score of the representation of $v$ in $\mathcal{A}_P$.

The \emph{\similarityrefiner{}} takes the inferred sinks
$\mathcal{S}_P$ and a set of precomputed embeddings of known sinks
$E_K$ as inputs. It implements a code-similarity based technique to
adjust the confidence scores of inferred sinks according to how
similar they are to known sinks, and returns the refined set of sinks
$\mathcal{S}_R$, which contains the same program elements as
$\mathcal{S}_P$, but with adjusted scores, and the embeddings for
predicted sinks $E_P$.

The role of the \emph{\feedbackrefiner{}} is to validate these
predictions computed over the test projects $\mathcal{D}_E$,
presenting the predicted sinks along with their confidence scores to
an analysis engineer to provide feedback about false positives, which
the \feedbackrefiner{} eliminates, leaving a final set of refined
sinks $\mathcal{S}_F$. This module uses the embeddings of predicted
sinks, $E_P$ provided by the \similarityrefiner{} to identify sinks
that may be similar to a false positive, allowing the engineer to
efficiently eliminate groups of similar false positives at once.

\subsection{\inferenceengine}\label{sec:adaptseldon}
\inferenceengine{} implements our Seldon-based approach for inferring likely
taint sinks by framing the problem as a linear optimization task.

\mypara{Computing program elements and triples}
Just like Seldon, we start by extracting triples of the form ($v_\textit{src}$,
$v_\textit{san}$, $v_\textit{snk}$) from the training projects $\mathcal{D}_T$,
where each triple denotes there is information flow from $v_\textit{src}$ to
$v_\textit{san}$ and from there to $v_\textit{snk}$, and the three program
elements are of the appropriate syntactic structure to potentially act as
source, sanitizer, and sink, respectively, as explained in
Section~\ref{sec:seldon}.

For capturing information flow between program elements, we use the standard
inter-procedural taint tracking framework of CodeQL instead of the points-to
analysis used by Seldon. The reason behind this choice is that we found that
building propagation graphs for JavaScript using Seldon's approach leads to both
spurious and missing flows.

For scalability reasons, we further restrict the set of sink candidates we build
triples for by focussing on candidates from the most popular libraries as
determined by the number of usages in JavaScript projects on LGTM.com.

\mypara{Program element representations}
We represent program elements using partial access paths~\cite{noregrets}, which
are a generalization of qualified names to the setting of JavaScript with its
highly dynamic object system and free use of higher-order functions. Access
paths are build from three basic operators: property access \code{p.q},
representing property \code{q} of the object represented by the base path
\code{p}; parameter access \code{p(i)}, representing the parameter \code{i} of
the function represented by the base path \code{p}; and result access
\code{p()}, representing the return value of the function represented by the
base path \code{p}.

For instance, the first argument of the invocation of \texttt{findByIdAndUpdate}
on line~\ref{line:eg1:updstart} of Figure~\ref{fig:motivating-example}(a) can
be represented by the following three access paths:

\begin{enumerate}
\item \texttt{findByIdAndUpdate(0)}, referring to it as the first argument to a
method called \texttt{findByIdAndUpdate};
\item \texttt{getquerySendResponse(0).*(0)}, referring to it as the first
argument to some method (the name being left unspecified) of an object that is
passed as the first argument to \texttt{getquerySendResponse};%
\footnote{This access path arises from a piece of code that is not shown in
Figure~\ref{fig:motivating-example}(a).}
\item \texttt{getquerySendResponse(0).findByIdAndUpdate(0)}, which is similar
but makes the name of the method concrete.
\end{enumerate}

Instead of Seldon's approach of allowing a program element to have multiple
representations, we only select one \emph{canonical representation} per program
element, which reduces the complexity of the constraint system. For canonical
representation we aim to find representations that are general enough to be
common across different projects, but still specific enough to capture semantic
differences. We choose the canonical representation by extracting features of
the representation such as its length and the number of occurrences of the
different kinds of accesses, and then assigning a score based on these features.
The weights for computing the score were determined semi-automatically by
computing representations of known sinks on a large set of training projects and
prioritizing common features.
In our example above, the first representation is chosen as canonical because it
provides enough information to obtain the program element, but disregards other
details in favor of generality. 

\mypara{Inferring new sinks specs using constraint solving}
Next, we construct a constraint system using the same approach as
Seldon and solve it using the CBC solver~\cite{forrest2005cbc}.
However, while Seldon combines the constraints extracted from
propagation graphs of all programs and builds a single optimization
objective to feed into the solver, we found this approach difficult to
scale. Instead, we solve constraints on a per-project basis, obtaining
one set of sink prediction specifications per project.  Then
we average prediction scores across all projects to obtain sink
prediction specifications $\mathcal{A}_P$.

\subsection{\getsinks{}}
Once we get the sink prediction specifications $\mathcal{A}_P$ we
proceed to predict concrete sinks by instantiating the specifications
on the test set $\mathcal{D}_E$. That is:
\begin{align*}
  \mathcal{S}_U = \{(v, p) | v \in \texttt{elements}(\mathcal{D}_E)
  \wedge(\texttt{repr}(v),\texttt{snk}, p) \in\mathcal{A}_P\\  \wedge
  (\texttt{repr}(v), \texttt{snk}, \_) \not\in \mathcal{A}_M \}
\end{align*}
Since our goal is to predict \emph{new} sinks, we furthermore remove
known sinks that are already modelled by the CodeQL library from this
set. The \getsinks{} module implements both these steps using CodeQL
library.

\subsection{\similarityrefiner}
Like Seldon, \inferenceengine{} assigns confidence scores based on
representation, so all program elements with the same canonical representation
are assigned the same score. However, sometimes representations can be too
coarse, representing both true sinks and false positives. One way to minimize
false positives is to use more specific representations. However, this can lead
to the opposite problem, where real sinks are missed. Further, it also impacts
scalability since it increases the number of constraint variables. The
\similarityrefiner{} tackles this problem by comparing the syntactic context of
predicted sinks with a corpus of known sinks collected across many projects
using code embeddings.

Our intuition behind using code similarity is that a sink candidate
which is used in a similar context as a known sink is more
likely to be true sink than one which is used in a different
context. Hence, we refine the predicted sinks by combining the
confidence scores computed by \inferenceengine{} and the
code-similarity based scores.

Algorithm~\ref{algo:similarity} describes our approach
conceptually. It takes the predicted sinks $\mathcal{S}_P$, and a set
of embeddings of known sinks $E_K$ indexed by representations. These
embeddings can be (but do not have to be) computed on the same set of
projects. The algorithm then computes, for each prediction, the
maximum similarity score with a known sink that has the same canonical
representation, using cosine similarity of code embeddings as
similarity metric as detailed in Algorithm~\ref{algo:embed}
below. This is done both for the enclosing statement
($\simm_\textit{stmt}$) and the enclosing function
($\simm_\textit{func}$). The final score of the sink is computed on
Line~\ref{line:sim:comb} as a combination of the original score and
the two similarity-based scores.

In practice, we pre-compute the set of embeddings, $E_K$, from a
  set of known sinks for performance gains. To obtain $E_K$, we build
  a training set comprising projects relevant to various queries (such
  as NoSqlInjection, XSS, and TaintedPath) and extract known sinks
  from them using CodeQL's pre-defined models. Our intuition here
  behind this approach is that the sinks that are currently unmodeled
  for a query may be used in syntactic contexts similar to not only
  the known sinks of that query but also to the ones in different
  queries as well.

\begin{algorithm}[!htb]
  \caption{Refining predictions using Code Similarity}
  \label{algo:similarity}
  \algosize{}
  \begin{flushleft}
    \hskip\algorithmicindent \textbf{Input}: Predicted sinks
    $\mathcal{S}_P$,  Known Sink Embeddings $E_K$ \\
    \hskip\algorithmicindent \textbf{Output}: Refined sinks $\mathcal{S}_R$
  \end{flushleft}  
  \begin{algorithmic}[1]
    \Procedure{\similarityrefinerFunc{}}{$\mathcal{S}_P,  E_K$}
    \State $\mathcal{S}_R = \emptyset$ \label{line:sim:eset}
    \State $E_P = \emptyset$ \label{line:sim:initep}
    \For{$(s, p) \in \mathcal{S}_P$} \label{line:sim:known}
    \State $\simm_\textit{stmt},
    \simm_\textit{func} = \textsc{ComputeSimilarityScore}(s, E_K)$
    \State $\mathcal{S}_R = \mathcal{S}_R \cup (s, (p +
    (\simm_\textit{stmt} +
    \simm_\textit{func})/2)/2)$ \label{line:sim:comb}
    \State $\text{stmt}_s , \text{func}_s = \texttt{getEnclosingCode}(s)$ \label{line:embed:enclosing1}
    \State $E_P(s) = (\texttt{Emb}(\text{stmt}_s) , \texttt{Emb}(\text{func}_s))  $ \label{line:sim:append}
    \EndFor 
    \State \Return $\mathcal{S}_R$ , $E_P$
    \EndProcedure
  \end{algorithmic}
\end{algorithm}
\vspace{-0.2in}

\Comment{
\begin{algorithm}[!htb]
  \caption{Refining predictions using Code Similarity}
  \label{algo:similarity}
  \algosize{}
  \begin{flushleft}
    \hskip\algorithmicindent    \textbf{Input}: Predicted sinks $\mathcal{S}_P$,  Known sinks $\mathcal{S}_K$ \\
    \hskip\algorithmicindent    \textbf{Output}: Refined sinks $\mathcal{S}_R$
  \end{flushleft}  
  \begin{algorithmic}[1]
    \Procedure{\similarityrefinerFunc{}}{$\mathcal{S}_P,  \mathcal{S}_K$}
    \State $\mathcal{S}_R = \emptyset$ \label{line:sim:eset}
     \For{$(s, p) \in \mathcal{S}_P$} \label{line:sim:known}
    \State $\boldsymbol{\simm}_\textit{stmt}, \boldsymbol{\simm}_\textit{func}
    = 0$
    \For{$(k, p') \in \mathcal{S}_K$} \label{line:sim:maxstart}
    \If{$\texttt{repr}(s) = \texttt{repr}(k)$}
    \State $\simm_\textit{stmt}, \simm_\textit{func} =
    \textsc{ComputeSimilarityScore}(s, k) $
    \State $\boldsymbol{\simm}_\textit{stmt} =  \texttt{max}(\boldsymbol{\simm}_\textit{stmt}, \simm_\textit{stmt})$
    \State $\boldsymbol{\simm}_\textit{func} =
    \texttt{max}(\boldsymbol{\simm}_\textit{func}, \simm_\textit{func})$    
    \EndIf    
    \EndFor \label{line:sim:maxend}
    \State $\mathcal{S}_R = \mathcal{S}_R \cup (s, (p +
    (\boldsymbol{\simm}_\textit{stmt} +
    \boldsymbol{\simm}_\textit{func})/2)/2)$ \label{line:sim:comb}
    \EndFor 
    \State \Return $\mathcal{S}_R$
    \EndProcedure
  \end{algorithmic}
\end{algorithm}
}

\mypara{Computing similarity using code embeddings}
Given a predicted sink, we compute its similarity to known sinks using
two kinds of code embeddings: 1) based on enclosing statement and 2)
based on enclosing function.
%
We use GraphCodeBERT~\cite{guo2020graphcodebert} to compute these
embeddings. GraphCodeBERT is a transformer-based model pre-trained
using large corpus of programs on a general task. GraphCodeBERT can be
fine-tuned to solve many downstream tasks in programming
languages. For our work, we use a publicly available pre-trained
GraphCodeBERT model used for clone
detection\footnote{\url{https://github.com/microsoft/CodeBERT/tree/master/GraphCodeBERT/clonedetection}}. Algorithm~\ref{algo:embed} describes our approach.

\begin{algorithm}
  \caption{Computing Max Similarity Score of a Sink}\label{algo:embed}
  \algosize{}
  \begin{flushleft}
    \hskip\algorithmicindent \textbf{Input}: Sink $s$, Embeddings $E_K$ \\
    \hskip\algorithmicindent \textbf{Output}: Statement similarity
    $\simm_\textit{stmt}$, Function similarity
    $\simm_\textit{func}$
  \end{flushleft}  
  \begin{algorithmic}[1]
    \Procedure{computeSimilarityScore}{$s, E_K$}
    \State $\text{stmt}_s , \text{func}_s = \texttt{getEnclosingCode}(s)$ \label{line:embed:enclosing1}
    \State $\text{embeddings} = E_K(\texttt{repr}(s))$ \label{line:embed:extractrep}
    \State $\simm_\text{stmt} = \texttt{max}\left\{\theta(\texttt{Emb}(\text{stmt}_s), e_\text{stmt})    |(e_{\text{stmt}}, \_) \in \text{embeddings}\right\} $  \label{line:embed:loopstart}
    \State $\simm_\text{func} = \texttt{max}\left\{\theta(\texttt{Emb}(\text{func}_s), e_\text{func})      |(\_, e_{\text{func}}) \in \text{embeddings}\right\}$ \label{line:embed:loopend}  
      \State \Return $\simm_\text{stmt}, \simm_\text{func}$  \label{line:embed:ret}
    \EndProcedure
  \end{algorithmic}
\end{algorithm}

The algorithm takes a sink program element $s$ and a embeddings map of
known sinks $E_K$ (indexed by representations) as inputs and returns
the maximum statement-based similarity $\simm_\textit{stmt}$ and
function-based similarity $\simm_\textit{func}$ for $s$ to any known
sink. The algorithm extracts the statement and function enclosing sink
$s$ (Line~\ref{line:embed:enclosing1}). Then, it obtains the
$\text{embeddings}$ for the representation of $s$
(Line~\ref{line:embed:extractrep}). Note that $\texttt{embeddings}$ is
a set of embeddings of sink program elements which have same
representation as sink $s$. Each element in this set is a tuple
containing the statement and function embeddings of a sink program
element.

The algorithm then computes the similarity of embedding of sink $s$ to
each embedding in the $\texttt{embeddings}$ set and stores the maximum
similarity scores in $\simm_\text{stmt}$ and
$\simm_\text{func}$
(Lines~\ref{line:embed:loopstart}-\ref{line:embed:loopend}). Here,
$\texttt{Emb}$ is a function which computes the embedding of a given
code-snippet (statement or function) using GraphCodeBERT. An embedding
is just a vector representation of a code snippet.  $\theta$ is a
function which computes the \emph{cosine similarity} of two input
vectors, i.e., the embeddings in this case.  Finally, it returns the
two maximum similarity scores (Line~\ref{line:embed:ret}).

\mypara{Combining the scores} Once, we compute the similarity
scores, we combine it with the confidence scores in
Algorithm~\ref{algo:similarity}, Line~\ref{line:sim:comb}. This step
improves the scores of predicted sinks which are similar to one or
more known sinks and penalizes the scores of predicted sinks which are
dissimilar.

\mypara{Computing similarity for sink predictions}
In addition to ranking the sinks prediction the \similarityrefiner{}
also computes embeddings $E_P$ for the sinks predictions
$\mathcal{S}_P$. Algorithm~\ref{algo:similarity} instantiates $E_P$ as
empty set (Line~\ref{line:sim:initep}) and updates the sink embeddings in each iteration (Line~\ref{line:sim:append}).

\subsection{\feedbackrefiner}
\begin{figure*}[!htb]
	\centering
        \vspace{-2em}	
     \begin{minipage}{.4\textwidth}
		\centering
		\includegraphics[width=0.8\linewidth]{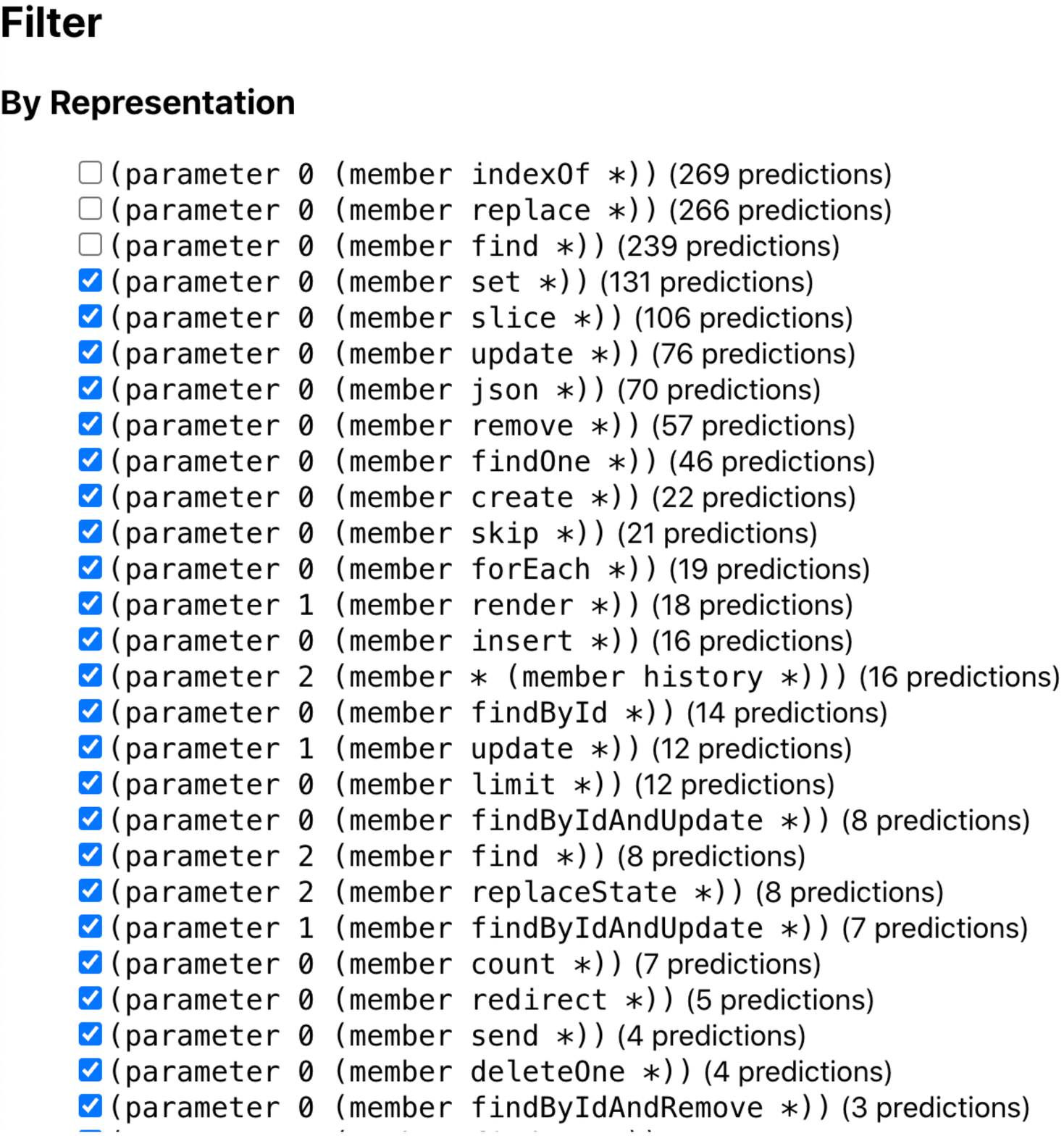}
	\end{minipage}%
	\begin{minipage}{0.6\textwidth}
		\centering
		\includegraphics[width=0.9\linewidth]{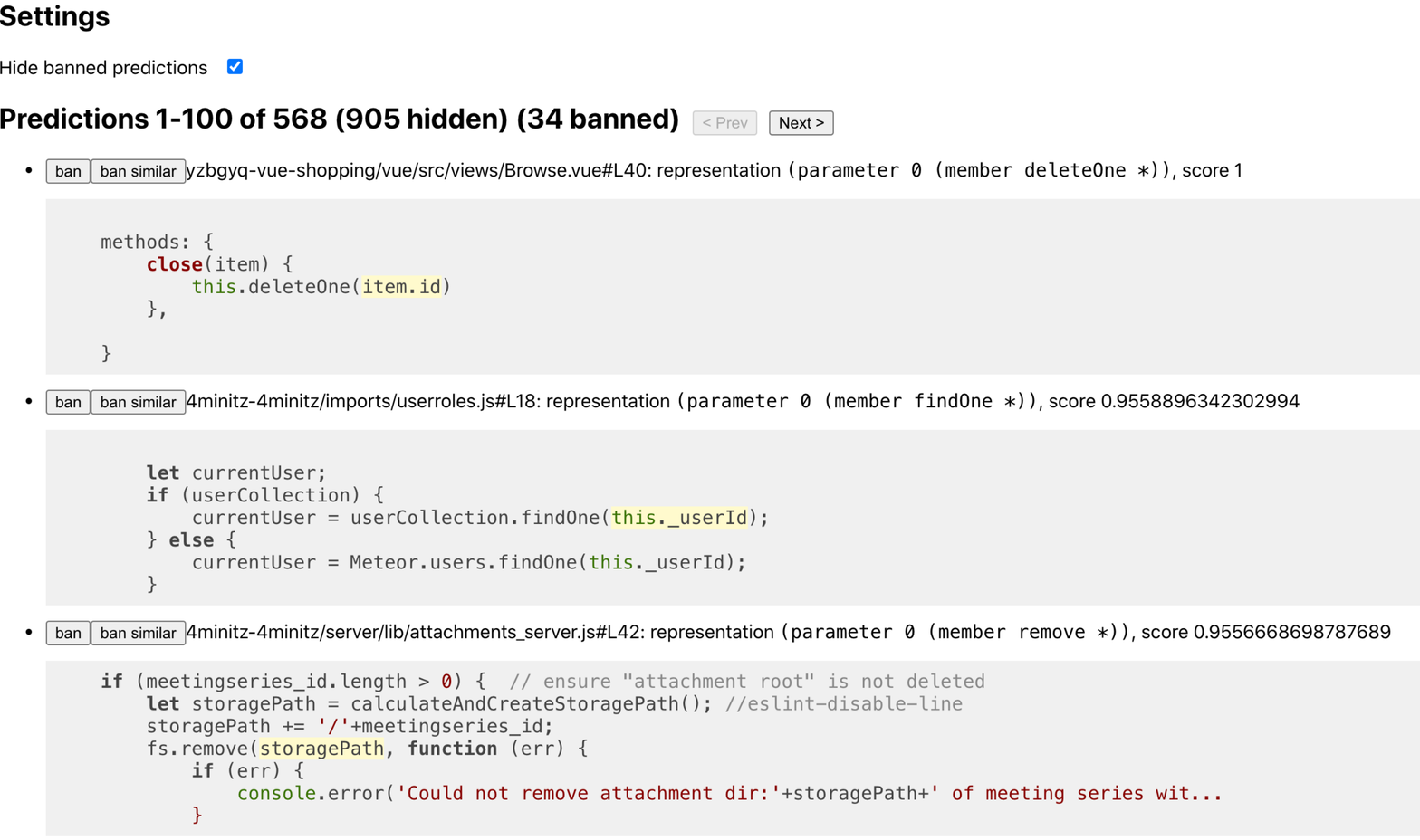}
		\label{fig:prob1_6_1}
	\end{minipage}
        \vspace{-2em}
	\caption{UX.  Left: selection of predictions by
          representation. Right: Sink candidates with score and option
          to ban individual or similar candidates.} \label{fig:UX}
        \vspace{-1em}
\end{figure*}

The \feedbackrefiner{} allows the developers to further refine the
predictions by incorporating their feedback. The \feedbackrefiner{}
provides a User Interface (UI) which displays all the predicted sinks
sorted by confidence scores.

Figure~\ref{fig:UX} shows a screenshot of the UI. It provides two
options with each prediction: ``ban'' -- which hides the corresponding
sink, and ``ban similar'' -- which hides the corresponding sink as
well as other sinks which are similar (up to some pre-selected
threshold).  The UI also shows the list of representations sorted by
the number of sinks each corresponds to.
Algorithm~\ref{algo:getsimilar} describes how we obtain the set of
sinks that are similar to given sink $s$. The algorithm computes the
similarity score of each sink $s'$ which has the same representation
as given sink $s$
(Lines~\ref{line:getsim:loopstart}-\ref{line:getsim:loopend}) using the
predicted sink embeddings $E_P$, computed by the
\similarityrefiner{}. It then selects sinks that have a similarity
score above user-defined threshold $\alpha$ and returns the selected
sinks to the caller.

The goal of our UI is to allow an experienced developer to quickly and
efficiently triage the list of predicted sinks. The developer can
remove individual (or similar) sinks that they consider to be false
positives. They can also filter the predictions by de-selecting
representations. For instance, a representation which matches too many
sinks may indicate that they are too coarse and mostly generate false
positives. The developer can easily hide the corresponding predictions
for such representations.

\begin{algorithm}[!htb]
  \caption{Filter similar sinks using Code Similarity}
  \label{algo:getsimilar}
  \algosize{}
  \begin{flushleft}
    \hskip\algorithmicindent \textbf{Input}:Sink $s$, Predicted sinks
    $\mathcal{S}_R$, Sink embeddings $E_P$, Similarity Threshold $\alpha$ \\
    \hskip\algorithmicindent \textbf{Output}: similar sinks $\mathcal{S}$
  \end{flushleft}  
  \begin{algorithmic}[1]
    \Function{GetSimilar}{$s$, $\mathcal{S}_R$, $E_P$, $\alpha$}
    \State $\mathcal{S} = \emptyset$ 
    \For{$(s', p) \in \mathcal{S}_R$} \label{line:getsim:loopstart}
    \If{$\texttt{repr}(s) = \texttt{repr}(s') \wedge s \ne s'$}
    \State $\text{e}_{\text{stmt}_s} , \text{e}_{\text{func}_s} =  E_P(s)$
    \State $\text{e}_{\text{stmt}_{s'}} , \text{e}_{\text{func}_{s'}} = E_P(s')$
    \State $\simm_\text{stmt} = \theta(\text{e}_{\text{stmt}_s},\text{e}_{\text{stmt}_{s'}})$  
    \State $\simm_\text{func} = \theta(\text{e}_{\text{func}_s}, \text{e}_{\text{func}_{s'}})$  
    \If  {$(\simm_\textit{stmt} +  \simm_\textit{func})/2 > \alpha $}
    \State $\mathcal{S}= \mathcal{S} \cup s'$
    \EndIf
    \EndIf
    \EndFor  \label{line:getsim:loopend}
    \State \Return $\mathcal{S}$
    \EndFunction
  \end{algorithmic}
\end{algorithm}





\section{Evaluation}\label{sec:evaluation}
To evaluate the practical usefulness of \tool{} we pose ourselves the following research questions:

\begin{itemize}
\item[\Q{1}:] Does \tool{} find new sinks that are as yet not covered by hand-written models?
\item[\Q{2}:] How much effort does it take to triage \tool{} results?
\item[\Q{3}:] How important are the different components of \tool{}?
\item[\Q{4}:] Can the predicted sinks be used to highlight new security alerts?
\end{itemize}

\mypara{Choosing JavaScript security queries} For all four research
questions, we focus on three representative CodeQL security queries
addressing some of the top 25 software vulnerabilities identified by
the MITRE CWE Top
25:\footnote{\url{https://cwe.mitre.org/top25/archive/2021/2021_cwe_top25.html}}
TaintedPath,\footnote{\url{https://git.io/JrRxW}} XSS,%
\footnote{\url{https://git.io/JrRAy}} and
NoSQLInjection.\footnote{\url{https://git.io/JrRNQ}} TaintedPath
detects path-traversal vulnerabilities where a potentially malicious
user can control the path of a file being read or written; XSS detects
client-side cross-site scripting vulnerabilities where potentially
malicious JavaScript code can be injected into the DOM of a web page;
and NoSQLInjection detects NoSQL-injection vulnerabilities where a
user can insert JavaScript code into a NoSQL query.

\begin{table*}[!thb]
  \caption{Results from manually labelling predicted
    sinks}\label{tab:manual-labelling}
  \tablesize{}
  \vspace{-0.1in}
    \begin{tabular}{l|r|r|r|r|r} \toprule
        \textbf{Query} & \textbf{\# Predictions} & \textbf{\# TPs} &
        \textbf{Min TP score} & \textbf{Coarsest TP repr} &
        \textbf{Max TP/FP similarity} \\ \midrule         
        TaintedPath & 4,611 & 56 & 0.58 & 3\% & 0.91 \\
        XSS & 10,504 & 436 & 0.75 & 7\% & 1.00 \\
        NoSQLInjection & 1,473 & 187 & 0.58 & 16\% & 0.93
        \\ \bottomrule 
    \end{tabular}
  \vspace{-0.1in}
\end{table*}

\mypara{Finding representative JavaScript projects per query} To
empirically evaluate the effectiveness of \tool{} and answer our
research questions, we need a corpus of JavaScript code to train our
model on and produce new predictions. For this purpose, we choose
open-source projects from GitHub.  While there is no shortage of such
projects, selecting projects at random would most likely have left us
with projects that do not use any APIs relevant to the three queries
we focus on. Instead, we choose projects where the existing CodeQL
query produces at least one alert (and hence the existing library
models identify at least one sink), the intuition being that these
projects perhaps also use other API, or as yet unmodeled parts of
APIs, relevant to the query.

To select candidate projects, we ran a query on all JavaScript
projects on LGTM.com~\cite{lgtm}, a cloud platform for running CodeQL
analysis results at scale on large numbers of open-source
repositories, in May 2021. Among the roughly 200,000 projects we
queried, we found 562 projects satisfying our criteria for
TaintedPath, 2834 for XSS, and 833 for NoSQLInjection.

We conducted two different experiments, one to address the first three
research questions, and the other to address the fourth question. We
will now describe the setup and outcomes of each experiment in turn,
and answer the research questions.

\subsection{Experiment 1: Manually labelling sink predictions}

For our first experiment, we used \tool{} to automatically identify sinks for
the three CodeQL queries, and then inspected the results.

\mypara{Experimental Setup} To keep the number of predictions
manageable, we randomly selected 100 projects per query, and then
split each set into a training set of 90 projects and a held-back test
set of 10 projects. We trained the \tool{} model on the training set
and produced predictions for the test set, filtering out any
previously known sinks for which CodeQL already has manually written
models. Finally, the fourth author (an experienced CodeQL analysis
engineer) manually labelled the predictions as true positives (that
is, sinks that are currently not modelled by the CodeQL standard
library but arguably should be), or false positives.

\mypara{Predictions} Table~\ref{tab:manual-labelling} presents the
results of this experiment. Each row presents the results of one
query. For each query, column~\textbf{\#Predictions} presents the total number of sink
predictions and column~\textbf{\#TPs} presents the number of true
positives.
This data allows us to answer \Q{1} in the affirmative: \emph{\tool{}
  does indeed find new sinks}.
We have reported missing sinks in CodeQL identified by \tool{} to the
CodeQL library maintainers on several occasions, which has already led
to numerous improvements to the manual models.%
\footnote{We contributed three pull requests, which have all been
  merged: \url{https://github.com/github/codeql/pull/5860},
  \url{https://github.com/github/codeql/pull/5262},
  \url{https://github.com/github/codeql/pull/4753}. Additionally, the
  library maintainers themselves implemented further improvements
  based on input from us:
  \url{https://github.com/github/codeql/pull/5862}.}

However, it is immediately obvious that the raw output of the ML model is too
noisy to be useful, with only a few percent of predictions being true positives.
This motivates the need for a tool like \tool{} to allow an analysis engineer to
efficiently triage the set of predictions and prune out false positives.

As described previously, \tool{} provides three metrics for
categorizing predictions: the \emph{score} of a prediction, the
\emph{coarseness} of its representation (that is, the percentage of
all predictions that have this representation), and the
\emph{similarity} of different predictions. The intuition is that a
prediction with a low score or high coarseness is likely to be a false
positive, and that false positives are likely to be similar to each
other, but not to true positives.

Table~\ref{tab:manual-labelling} shows some statistics that allow us
to test this claim. Column \textbf{Min TP score} presents the
minimum score of a true positive, which is above 0.5 for each query;
this suggests that predictions with a score below 0.5 can be
disregarded in practice. Column \textbf{Coarsest TP repr}
presents the maximum coarseness of a true positive, that is, the
percentage of predictions that have the same representation as a true
positive. This value varies quite a bit between queries, from 3\% for
TaintedPath to 16\% for NoSQLInjection.  Disregarding predictions
whose representation accounts for more than 20\% of all predictions
seems like a reasonably safe thing to do in practice, but the evidence
is not clear cut in this case.

\begin{table*}[!thb]
  \caption{Metrics for triaging effort, with similarity threshold 0.95
    and coarseness threshold 20\%}\label{tab:triaging-effort}
  \tablesize{}
  \vspace{-0.1in}
    \begin{tabular}{l|r|r|r|r}\toprule 
        \textbf{Query} & \textbf{\# Discarded (Due to Score + Coarseness)} & \textbf{\# Remaining FPs} & \textbf{\# Steps to Triage} & \textbf{False Negatives} \\
        \midrule 
        TaintedPath & 3,007 (1,025 + 1,982) & 1,548 & 523 & 0 \\
        XSS & 2,136 (2,136 + 0) & 7,932 & 2,874 & 10 \\
        NoSQLInjection & 666 (666 + 0) & 620 & 243 & 0 \\ \bottomrule
    \end{tabular}
\vspace{-0.1in}
\end{table*}

Finally, column \textbf{Max TP/FP similarity} shows the maximum
similarity between a true positive and a false positive. Recall that
our prototype allows a user to dismiss not just a single false
positive they have identified, but also all other predictions that are
\emph{sufficiently similar} to it. Here, ``sufficiently similar''
should be chosen in such a way that it is unlikely that any of the
predictions dismissed alongside the false positive are true
positives. Unfortunately our experiment shows that this is not
achievable with our current similarity metric: for XSS, there are true
positives that are indistinguishable from false positives in terms of
code similarity, meaning their similarity score is 1. For the other
queries, a similarity score of 0.95 looks to be a safe cut-off.


\mypara{Triaging effort} To estimate the effort required to triage the
set of predictions, we count the number of predictions that are
discarded due to not meeting the cut-off for score or coarseness, and
the number of steps that would be required to triage the remaining
predictions, as well as the number of true-positive predictions that
would be wrongly discarded during this process.  These are, of course,
best-case estimates since we are using cut-offs established on the
same dataset.

Table~\ref{tab:triaging-effort} presents the results of this computation.
Column~\textbf{\# Discarded} shows the number of predictions that are discarded
(with details on how many were discarded due to low score and high coarseness,
respectively, in brackets); column~\textbf{\# Remaining FPs} shows the number of
false-positive predictions that are not discarded; column~\textbf{\# Steps to
Triage} shows the number of steps needed to triage the remaining predictions;
and column~\textbf{\# False Negatives} the number of true positives that are
missed in this process. We can see that score and coarseness act as a very
useful first filter, discarding 65\% of predictions for TaintedPath, 20\% for
XSS, and 45\% for NoSQLInjection. After that, the analysis developer still needs
to identify and dismiss the remaining false positives, but as the table shows
the similarity-based multi-dismissal feature significantly reduces that effort,
which each step on average dismissing about three false positives in one go.

For XSS, multi-dismissal results in ten false negatives, since, as we discussed
above, there are true positives that are indistinguishable from false positives
in terms of code similarity. For the other queries, the number of false
negatives is zero.

\begin{figure}[!htb]
  \centering
  \begin{tabular}{c}
    \includegraphics[width=0.35\textwidth]{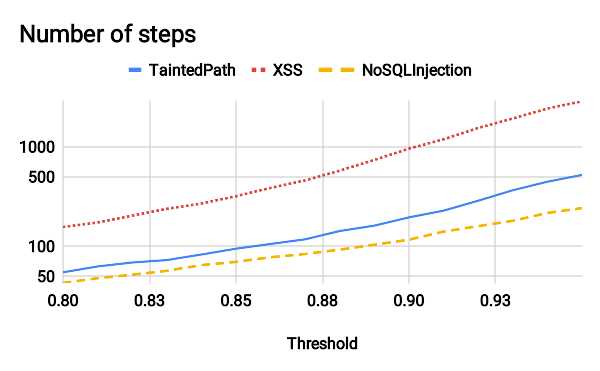} \\
    \includegraphics[width=0.35\textwidth]{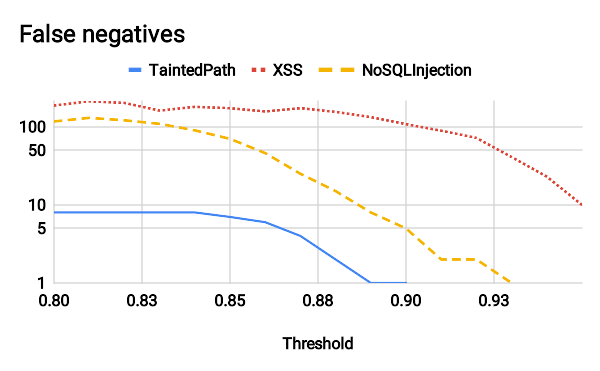} 
  \end{tabular}
    \vspace{-0.2in}
  \caption{Impact of similarity threshold on triaging effort
    (top) and false negatives (bottom); y-axis is log scale.}
  \label{fig:triaging-effort}
  \vspace{-0.2in}
\end{figure}

Figure~\ref{fig:triaging-effort} shows how the number of steps and the number of
false positives vary with the similarity threshold: for each of our three
queries, we compute both metrics for similarity thresholds between 0.80 and
0.95. As expected, decreasing the similarity threshold makes triaging faster:
with a threshold of 0.80, the number of steps is 55 for TaintedPath, 157 for
XSS, and 43 for NoSQLInjection, which is about 1/10th to 1/20th the number
for a threshold of 0.95 as shown in Table~\ref{tab:triaging-effort}. Of course,
this comes at the price of missing true positives: at similarity threshold 0.80,
TaintedPath misses eight true positives, XSS 187, and NoSQLInjection 117, which is
significantly above the values for 0.95 shown in Table~\ref{tab:triaging-effort}.

Our answer to \Q{2} is, therefore, nuanced: an analysis engineer using
\tool{} needs to be aware that the vast majority of predictions are
likely to be false positives, but the various metrics provided by
\tool{} can be used to trade off triaging effort against them.

\mypara{Importance of \tool{} components} The novelty of \tool{} lies in
combining the triple-mining approach of Seldon with a code-similarity metric to
weed out false positives. A natural question is whether this combination works
better than either component alone, which we investigate now.

On one hand, we might conceivably do away with the triple computation
altogether, and simply consider all data-flow nodes that could
potentially be sinks, relying entirely on code similarity to rank and
triage them. However, this is not a viable approach: the number of
potential sinks across our ten test projects is 992,035 for
TaintedPath, 1,308,150 for XSS, and 105,651 for
NoSQLInjection. Overall, this means that the triple-computation step
reduces the number of predictions by about two orders of magnitude.

On the other hand, we could discard the code-similarity step, but, as
already discussed, Table~\ref{tab:triaging-effort} shows that without
similarity-based multi-dismissal the triaging effort would be about
three times as big.  In summary, then, our answer to \Q{3} is that
both components of our approach make their own important contribution
towards easing the reviewing burden.

\subsection{Experiment 2: Analyzing security alerts}
To answer \Q{4}, we take our three queries and boost them, that is, we use
\tool{} to predict new sinks on a training set of projects, and then include
them among the set of sinks recognized by the query to yield a boosted query. We
then run that boosted query on a test set of projects, and consider the new
alerts it produces on these projects (compared to the original query), and
evaluate whether they are correct.

\mypara{Experimental Setup} Manually evaluating whether new alerts are
correct is labor-intensive, of course, so we use an alternative
strategy to evaluate \tool{}'s performance: for each given CodeQL
query $Q$, we obtain an older version $Q_\text{old}$ of the same query
from the version history of CodeQL. This query will have the role of
$\mathcal{A}_M$ and we boost it with \tool{} it produce
$\mathcal{A}^{boosted}_M$. We call this query
$Q_\text{old}^\text{boosted}$.
We analyze the boosted query on a set of test projects to generate
alerts, and compare them against the alerts generated by
$Q_\text{new}$ of $Q$ on the same set of projects, which we use as
ground truth.

%
%
%

\begin{table*}[!htb]
  \caption{Results from comparing old versions of queries
    boosted with \tool{} to the latest version of the same
    query. Averaged over three runs on 200 projects, with random
    50-50 splits to obtain test and training sets in each
    round.}\label{tab:atm-eval}
  \tablesize{}
  \vspace{-0.1in}
	\begin{tabular}{l|r|r|r|r} \toprule 
		\textbf{Query} 
		& \textbf{Alerts to Recover} 
		& \textbf{Alerts Recovered} 
		& \textbf{Spurious Alerts} 
		& \textbf{Projects with Alerts to Recover} \\
		\midrule 
		TaintedPath   & 58.33 &  46 & 1909 & 19\\ 
		XSS   & 15  & 14  & 406 &  4.33\\ 
		NoSQLInjection  & 303  & 266.67&  719.67 &  38.67 \\ 
                \bottomrule
	\end{tabular}

\vspace{-0.1in}
\end{table*}

To run this experiment, we selected 200 different projects for each of
the three queries we consider. Then, for each query we run three
rounds of the boosting process described above, randomly splitting the
projects into 100 projects for training and 100 for testing in each
round.

\mypara{Results} In Table~\ref{tab:atm-eval} we present the results
averaged over the three rounds for each query. Column~\textbf{Alerts
  to Recover} shows the average number of new alerts produced by
$Q_\text{new}$ that are not in the original query
$Q_\text{old}$. Column~\textbf{Alerts Recovered} shows how many of
these new alerts are also flagged by the boosted query
$Q_\text{old}^\text{boosted}$ on average.  Conversely,
Column~\textbf{Spurious Alerts} shows how many of the new alerts from
the boosted query are \emph{not} flagged by $Q_\text{new}$. We
consider them as false positives, even though it is possible that some
of them are actual true positives not captured by $Q_\text{new}$.
Finally, column~\textbf{Projects with Alerts to Recover} shows how
many of the 100 test projects had any alerts to recover on
average.

In response to \Q{4}, we can say that \tool{} succeeds in predicting
sinks that lead to security alerts, and its recall with respect to new
query versions is high. The false positive rate is also quite high,
however, which agreess with Experiment~1 results. It is worth
noting that in this experiment we do not filter out predictions with
very coarse representations, which may exacerbate this problem, and of
course (as noted above) our labelling of false positives is
over-approximate, so the actual number of false positives is lower.

\subsection{Threats to validity}
The main threat to the validity of the results from Experiment~1 is bias in the
manual labelling. To counter this threat, we randomly selected 20 predictions
for each query and gave them to CodeQL experts not involved in this project to
label, and compared the results with our own labelling. For TaintedPath we
agreed on all 20 predictions, for XSS on 17 (with the external expert marking
three predictions as true positives that we had dismissed as false positives),
and for NoSQLInjection again on all 20. These results give us some amount of
confidence in the reliability of our labelling, perhaps suggesting a slight bias
towards dismissing predictions as false positives on part of the fourth author.

The small number of queries and of projects investigated in both experiments
also puts a limit on the quantitative generalizability of our results. For the
time being, we content ourselves with qualitative conclusions: \tool{} finds
additional sinks missed by manual modeling, but incurs a substantial number of
false positives; the techniques it offers for organizing predictions reduce the
effort required to prune them, however, and the predicted sinks are useful in
finding security alerts.

\section{Related Work}\label{sec:related-work}
\mypara{Taint Specification Mining} There are several prior approaches
for inferring information flow specifications from
programs. 
Merlin~\cite{livshits2009merlin} models information flow paths in C\#
programs using probabilistic constraints and solves them using factor
graphs. However, Merlin only works on statically typed languages
(C\#). Further, inference using factor graphs is much less scalable
than approaches using linear constraints (which both our work and
Seldon~\cite{seldon} uses). Seldon~\cite{seldon} was originally
evaluated on Python programs. In this work we adapt their approach for
JavaScript programs and improve on their technique by incorporating
code similarity-based filtering mechanism and refinement of
predictions using user feedback. SUSI~\cite{rasthofer2014machine} is a
SVM-based approach for detecting sources and sinks in Android
APIs. However, their approach relies on static program features and
similarity of APIs with similar signatures which are hard to obtain
for dynamic languages like JavaScript. Staicu
\etal{}~\cite{staicu2020extracting} use dynamic analysis for detecting
taint specifications for JavaScript. However, their method depends on
extracting information by executing existing test-suites. This
approach may miss sources/sinks which are not covered by the
test-suite. In contrast, \mbox{\tool{} is more likely to over-approximate
true sources/sinks.}

\mypara{Taint Analysis} There are several
static~\cite{yang2012leakminer,arzt2014flowdroid} and
dynamic~\cite{clause2007dytan,wei2013practical} taint analyses
proposed in literature and employed for detecting security issues or
other vulnerabilities in code. \tool{} can aid existing taint analysis
techniques by filling in the gap of missing taint specifications and
improve their effectiveness.

\mypara{Feedback driven analyses} Raghothanam et
al.~\cite{raghothaman2018user} leverage user-feedback to improve an
underlying probabilistic static analysis. In contrast, our
user-feedback leverages code-similarity as a postprocessing step to
help triage warnings.

\mypara{Learning Based Approaches for
  Predicting Program Properties}
GraphCodeBERT~\cite{guo2020graphcodebert} is a transformer-based
approach for learning semantic information from code. We adopt their
approach for improving the precision of \tool{} by identifying similar
events which are more likely to have similar roles (e.g., sinks).
JSNice~\cite{raychev2015predicting} is another learning-based approach
for predicting syntactic or semantic program properties 
for JavaScript.
Typilus~\cite{allamanis2020typilus} is Graph Neural Network (GNN)
based approach for predicting variables types for Python. As such,
these techniques may also be leveraged by future approaches to improve
the precision of \tool{}'s results.

\section{Conclusion}\label{sec:conclusion}
In this paper, we described our experience combining machine-learning
based taint specification inference of sinks along with manual
modelling for important CodeQL security queries for JavaScript. We
also describe how we leverage code-similarity metrics and
user-feedback to help analysis engineers effectively triage the
predictions to prune spurious predictions.

In future work, we are working to extend \tool{} to infer source and
sanitizer specifications, as well as \emph{taint-flow} and aliasing
specifications. We are also working on incorporating approaches based
on {\it abductive inference} of library
specifications~\cite{zhu-aplas13}.
\dg{What about extending the UX to also promote sinks similar to true positives. Use the refined sinks for finding alerts?}

\section*{Acknowledgments}
We want to thank Ian Wright, Henry Mercer, Oege de Moor  and Madanlal Musuvathi for supporting this project.

\balance

\bibliography{references}

\end{document}